\documentclass[aps,pra,showpacs,preprintnumbers,amsmath,amssymb,footinbib]{revtex4}
\usepackage{graphicx,epsfig}
\usepackage{bm}
\usepackage{etoolbox}
\usepackage{dcolumn}
\usepackage{xcolor}
\usepackage[breaklinks=true,colorlinks,citecolor=blue,linkcolor=blue,urlcolor=blue]{hyperref}

\def\noi{\noindent}
\def\bc{\begin{center}}
\def\ec{\end{center}}
\newcommand{\bea}{\begin{equation}}
\newcommand{\eea}{\end{equation}\noi}
\newcommand{\ber}{\begin{eqnarray}}
\newcommand{\eer}{\end{eqnarray}\noi}
\usepackage{graphicx}
\usepackage{amsmath}

\begin{document}

\title{Prospects for Lorentz Invariance Violation searches with top pair production at the LHC and future hadron colliders}

\author{Aur\'elien Carle\thanks{e-mail: a.carle@ipnl.in2p3.fr},
        Nicolas Chanon\thanks{e-mail: n.chanon@ipnl.in2p3.fr2}, 
        St\'ephane Perri\`es\thanks{e-mail: s.perries@ipnl.in2p3.fr} 
}
\affiliation{Institut de Physique Nucl\'eaire de Lyon, Universit\'e de Lyon, Universit\'e Claude Bernard Lyon 1, CNRS-IN2P3, Villeurbanne 69622, France}

\begin{abstract}

This paper investigates a signature of Lorentz invariance violation with the $t\bar{t}$ production at the LHC and future hadron colliders. Possible deviations from the Lorentz symmetry remain poorly constrained in the top quark sector. With a dedicated analysis of $t\bar{t}$ events produced at the LHC, bounds in the top sector can be improved by up to three orders of magnitude relative to Tevatron \cite{Abazov:2012iu}, the only measurement existing so far. The sensitivity will be even further enhanced at the HL-LHC and future colliders. 
\end{abstract}

\maketitle

\section{Introduction}
\label{intro}

Lorentz Invariance is a fundamental symmetry of the Standard Model (SM), however it is not expected to be conserved necessarily at the high energy scale of quantum gravity (e.g. in string theory \cite{Kostelecky:1988zi} or quantum loop gravity \cite{Gambini:1998it}) where spacetime could undergo violent fluctuations. 
Quantum field theories with non-commutative geometries 
introduce a fundamental length-scale, hence exhibiting Lorentz Invariance violation (LIV) \cite{Carroll:2001ws}. 
Cosmologies with spacetime varying couplings are natural in some grand unified theories and lead to signatures of LIV  \cite{Kostelecky:2002ca}. 
Remnants from the symmetry breaking would manifest themselves at a lower energy, and constitute an appealing signature.

Such signatures are predicted within the “Standard Model Extension” (SME) \cite{Colladay:1996iz}\cite{Colladay:1998fq}, an Effective Field Theory (EFT) considering all possible Lorentz- and CPT-violating operators in the Lagrangian (CPT breaking implies Lorentz violation for local theories \cite{Greenberg:2002uu}) in a model-independent way, preserving gauge invariance, renormalizability, locality and observer causality. 
The SME was tested with atomic clocks, penning traps, matter and antimatter spectroscopy, colliders and astroparticle experiments (for a review, see \cite{Liberati:2013xla}), and an impressive set of results was compiled \cite{Kostelecky:2008ts}. 

At hadron colliders the quark sector can be probed. The quark sector is constrained mostly with flavour measurements from neutral meson mixing. The most recent search for LIV and CPT breaking in the b-quark sector was performed recently at LHCb, using changes in $B_{(s)}$ mixing with sidereal time \cite{Aaij:2016mos}. Within the SME, values of the coefficients can be species-dependent and need not  be the same for each quark flavour. 
The top quark sector remains a vastly unexplored area for LIV searches, with only one actual measurement ever performed. The D\O~experiment at Tevatron measured Lorentz violating (and CPT-conserving) SME coefficients in the top quark sector \cite{Abazov:2012iu}, and found no evidence for LIV with 10\% absolute uncertainty. 
The LHC is a top factory, producing top quark pairs ($t\bar{t}$) with a high cross section, and provides a unique opportunity for measuring precisely SME coefficients in the top sector. 
In this paper, we will derive the expected sensitivity to SME coefficients using the top pair production signature.

\section{Theoretical setup}
\label{sec:1}

The SME describes the interaction of Lorentz-violating ``background fields'' with the SM particles \cite{Colladay:1998fq}. They can arise in theories like the string scenario \cite{Kostelecky:1988zi}, where certain fields acquire a non-zero vacuum expectation value thereby spontaneously breaking the Lorentz symmetry.
Within the SME, the EFT Wilson coefficients are identified with such vacuum expectation values and are constant in a given inertial frame, taken by convention to be the sun-centered frame \cite{Bluhm:2001rw}. The sun-centered frame can be considered as inertial in the lifetime of a physics experiment. The origin is placed at the sun center, the Z-axis directed north and parallel to the earth rotation axis, the X-axis is pointing to the vernal equinox of year 2000 in the celestial sphere, while X- and Y-axis are defining the equatorial plane, lying at an angle of $\approx23^{\circ} $ relative to the ecliptic.

In this paper, we are interested in the Lorentz violating CPT-even part of the Lagrangian density modifying the top quark kinematics \cite{Berger:2015yha}:
\begin{equation}
\label{LIVLagrangianDirac}
\mathcal{L} \supset \frac{1}{2}i(c_L)_{\mu\nu} \bar{Q}_t \gamma^{\mu} \overleftrightarrow{D}^{\nu} Q_t 
+ \frac{1}{2}i(c_R)_{\mu\nu} \bar{U}_t \gamma^{\mu} \overleftrightarrow{D}^{\nu} U_t 
\end{equation}
where $(c_L)_{\mu\nu}$ and $(c_R)_{\mu\nu}$ are 4$\times$4 matrices containing top quark SME coefficients (constant in the sun-centered rest frame), $Q_t$ is the third generation left-handed quark doublet, $U_t$ is the right-handed charge-2/3 top singlet, and $D^\nu$ is the gauge-covariant derivative.

A laboratory frame on earth moves around the earth rotation axis, thus the matrices $c_{\mu\nu}$ are oscillating within this frame during a sidereal day.
Top quark interactions with $c_{\mu\nu}$ result in a distinctive signature: the cross section for $t\bar{t}$ production is modulating with sidereal time in the frame of the experiment, thus exhibiting Lorentz violation. The first dedicated search for such signature in the top sector was performed by  D\O~\cite{Abazov:2012iu}.

\section{Top pair production in the SME}
\label{sec:2}

The matrix elements for $t\bar{t}$ production in the SME were calculated analytically \cite{Berger:2015yha} at leading order in perturbative QCD, assuming narrow-width approximation. Under the hypothesis that the parton distribution functions in the proton are not modified (which is the case if only the top quark receives non-zero SME coefficients), and since the phase space expression stays identical (neglecting second order modification of the dispersion relation), the ratio of SME over SM cross section is:
\begin{equation}
\label{eqFt}
w = \frac{|\mathcal{M}_{SME}|^2}{|\mathcal{M}_{SM}|^2} 
\end{equation}
with $\mathcal{M}_{SME}$ and $\mathcal{M}_{SM}$ the matrix elements for $t\bar{t}$ production in the SME and in the SM. In the laboratory frame, the ratio is expressed as $w(t) = 1 + f(t)$, with:
\begin{equation}
\label{eqFt2}
\begin{split}
f(t) = (c_{L,\mu\nu}+c_{R,\mu\nu}) R^{\mu}_{\alpha}(t) R^{\nu}_{\beta}(t) \Big(\frac{\delta_p P}{P} + \frac{\delta_v P}{P} \Big) ^{\alpha\beta}\\
+ c_{L,\mu\nu} R^{\mu}_{\alpha}(t) R^{\nu}_{\beta}(t) \Big( \frac{\delta F}{F} + \frac{\delta \bar{F}}{\bar{F}} \Big)^{\alpha\beta}
\end{split}
\end{equation}
where $P$ is the SM matrix element squared for $t\bar{t}$ production (either quark-antiquark annihilation or gluon fusion), $F$ and $\bar{F}$ are the SM matrix elements for top and antitop decay, while $\delta_p P$, $\delta_v P$, $\delta F$, $\delta \bar{F}$ are the SME modifications in the matrix element due respectively to propagator, production vertex, and in the top and antitop decay. 

The rotation matrix $R(t)$ implements the change of reference frame from the laboratory frame to the sun-centered canonical, and depends on the sidereal time, owing to the earth rotation around its axis with an angular velocity $\Omega=7.29 \times 10^{-5} rad \cdot s^{-1}(SI)$ (the earth boost due to its revolution around the sun is negligible relative to the top quark boost produced in collisions). 
In the following developments, for definiteness we will consider the rotation matrix constructed with the CMS experiment \cite{Chatrchyan:2008aa} as laboratory frame. CMS is located at an azimuth of approximately $\theta = 101.28 ^{\circ}$ on the LHC ring; the latitude of the CMS interaction point is $\lambda = 46.31^{\circ}$, and the longitude is $\ell = 6.08 ^{\circ} E$. Because the ATLAS experiment \cite{Aad:2008zzm} is located at the opposite azimuth on the LHC ring, both experiments would lead to similar results in the following studies.

Samples of $t\bar{t}$ with dilepton decay $t\bar{t} \rightarrow b e^{\pm} \nu \bar{b} \mu^{\mp} \bar{\nu}$ are generated with $\mbox{\textsc{MadGraph}-aMC@NLO}$ 2.6 \cite{Alwall:2014hca} at leading order. 
The ratio $w$ can be considered as an event weight, to be applied to simulation events generated at leading order in QCD. Each simulated event is given a weight $w$, depending on the event kinematics and on the sidereal timestamp (attributed to the event according to its event number). The selection criteria required on reconstructed particles are taken from \cite{Khachatryan:2016kzg}. 
Two jets are selected, arising from b-quark hadronization, with transverse momenta $p_T > 30$ GeV and pseudorapidity $|\eta|<2.4$. Two leptons are required to have $p_T>20$ GeV and $|\eta|<2.4$. No requirement on missing transverse momentum is imposed, instead the selection on the invariant mass $m_{e\mu}>20$ GeV is applied to reject Drell-Yan background of $\tau$ lepton pairs with low invariant mass. 
The $t\bar{t}$ dilepton channel \cite{Aaboud:2016pbd} provides a sensitivity similar to the sensitivity of the lepton+jet channel \cite{Khachatryan:2016kzg}, that was used in the D\O~analysis~\cite{Abazov:2012iu}.

\section{Anatomy of the LIV signatures in $t\bar{t}$}

The function $f(t)$ is computed in $t\bar{t}$ simulated events. In eq.~\ref{eqFt2}, we average terms relative to the event kinematics (that do not depend on time): $<A_P^{\alpha\beta}> = <( \frac{\delta_p P}{P} + \frac{\delta_v P}{P} )^{\alpha\beta}>$ and $<A_F^{\alpha\beta}> = <(\frac{\delta F}{F} + \frac{\delta \bar{F}}{\bar{F}})^{\alpha\beta}>$. 
Off-diagonal elements in the matrices $A_P$ and $A_F$ are much smaller than the in-diagonal elements, and are neglected in calculating the sinusoidal functions $f(t)$. 

Four benchmark scenarios of SME coefficients, taken from \cite{Abazov:2012iu}, are studied: \\
1) $c_{L,\mu\nu}\neq 0$ while $c_{R,\mu\nu}=0$, \\
2) $c_{R,\mu\nu} \neq 0$ while $c_{L,\mu\nu}=0$, \\
3) $d_{\mu\nu} =  (c_{L,\mu\nu}-c_{R,\mu\nu})/2 \neq 0$ while $c_{\mu\nu} = (c_{L,\mu\nu}+c_{R,\mu\nu})/2=0$.\\
4) $c_{\mu\nu} \neq 0$ while $d_{\mu\nu} = 0$.

The matrices $c_{\mu\nu}$ ($\mu$ or $\nu = T,X,Y,Z$) are assumed to be symmetric (the antisymmetric part can be absorbed in other SME terms in the Lagrangian) and traceless (the trace is Lorentz invariant).
Coefficients of the type $c^{TT}$ impact only the total $t\bar{t}$ cross section \cite{Berger:2015yha} and are not considered further, since there is no handle to extract them in genuine $t\bar{t}$ production.
Similarly, $c^{ZZ}$ coefficients are not studied here, since by construction there is no sensitivity induced by earth rotation in the direction transverse to the equatorial plane. 
As a consequence, there is no sensitivity to $c^{TZ}=c^{ZT}$ coefficient either. 
Eventually, $c^{TX}=c^{XT}$ and $c^{TY}=c^{YT}$ could be measured, but according matrix elements contributing in $A_P$ and $A_F$ are found to be negligible and these coefficients are not considered further. 

\begin{figure}[h!]
        \begin{center}
            \label{ampl:1}
            \includegraphics[scale=0.4]{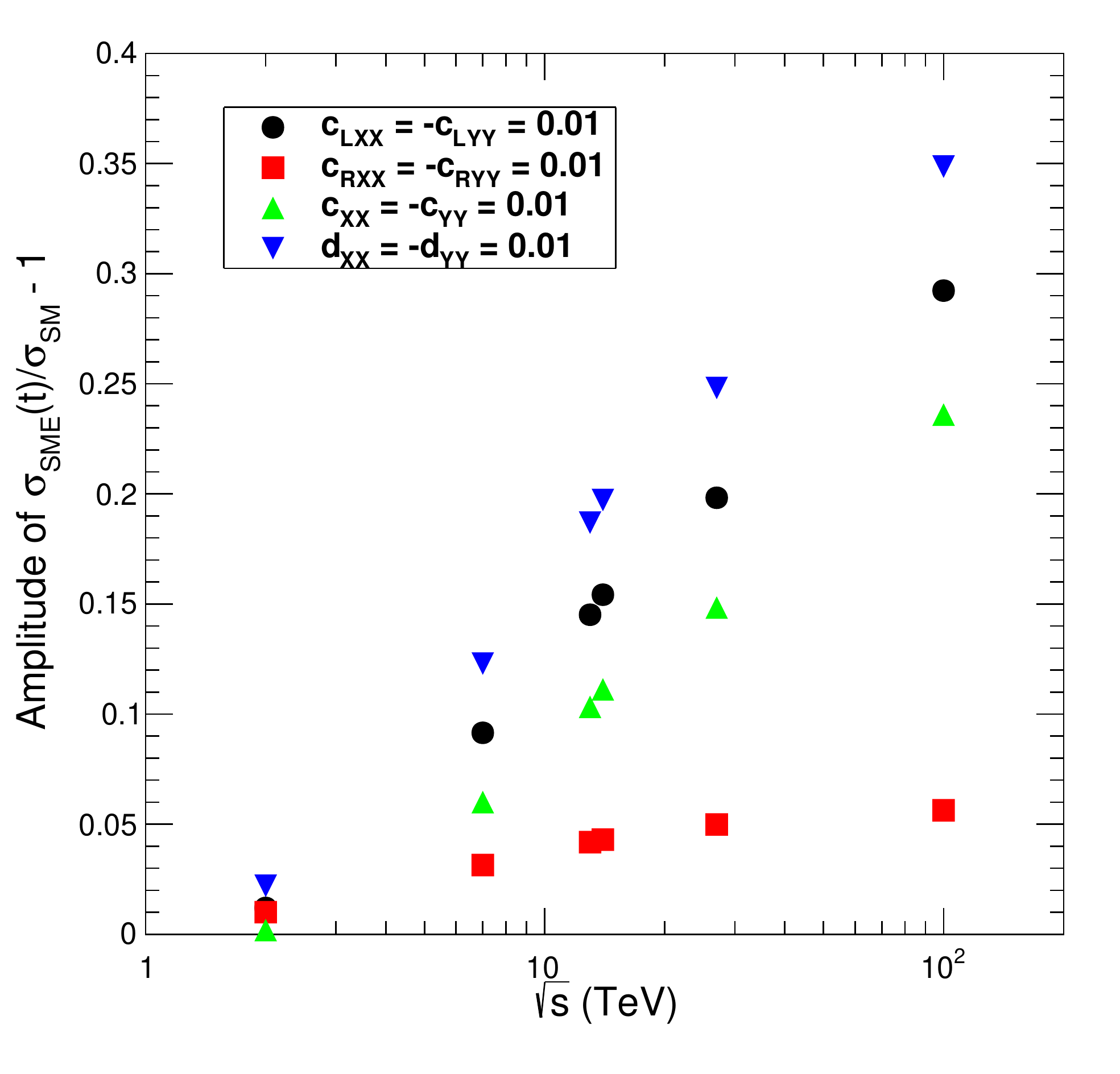}
            \includegraphics[scale=0.4]{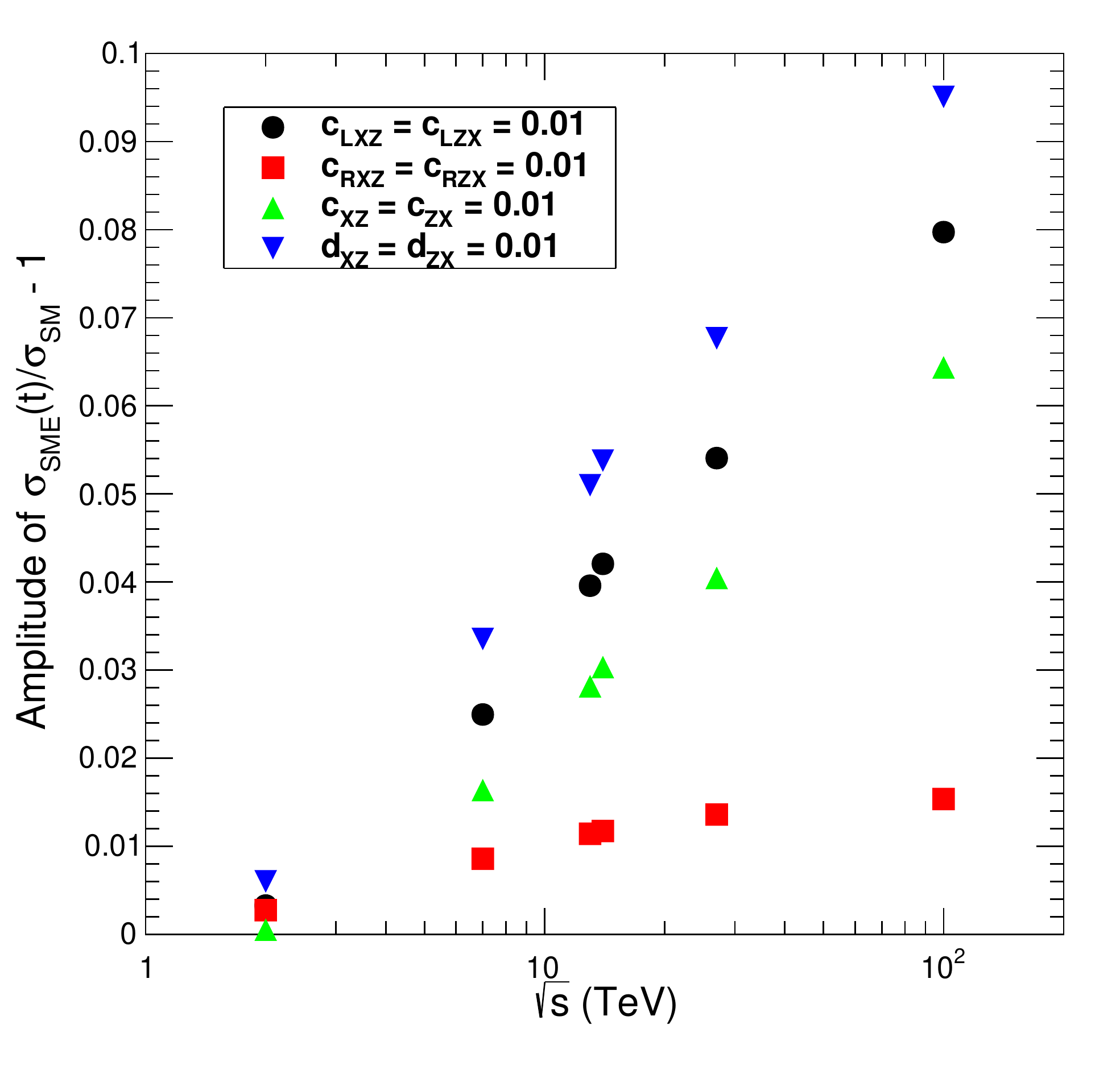}
            \caption{\label{fig:1}Amplitude of $f(t)=\sigma_{SME}/\sigma_{SM}-1$ in p--p collisions at center-of-mass energies corresponding to Tevatron, LHC, HL-LHC, HE-LHC and FCC energies, using CMS location as laboratory frame, for $c_{\mu \nu}$ benchmarks.}
        \end{center}
    \end{figure}

The analysis will focus on the sinusoidal signals expected for $c_{XZ} =c_{ZX} \neq 0$ and $c_{YZ} =c_{ZY} \neq 0$ with harmonics at a period of one sidereal day; $c_{XX} =-c_{YY} \neq 0$ and $c_{XY} =c_{YX} \neq 0$ with a period of half a sidereal day. 
Amplitudes of the $f(t)$ functions, at selected center-of-mass energies in p--p collisions, are shown in Fig~\ref{fig:1}. 
Amplitudes of $f(t)$ are found to be the same in the scenarios $c_{XY} =c_{YX}$ and $c_{XX} =-c_{YY}$ on the one hand, as well as in the scenarios $c_{XZ} =c_{ZX}$ and $c_{YZ} =-c_{ZY}$ on the other hand (although phases of the sinusoidal functions are different). Larger amplitudes of $f(t)$ are found in the benchmark scenarios $c_{XY} = c_{YX}$ and $c_{XX} = -c_{YY}$: this confirms that the experiment have higher sensitivity to $c_{\mu\nu}$ components along directions purely in the equatorial plane. 

It is observed on Fig~\ref{fig:1} that the amplitude of $f(t)$ function is growing with $\sqrt{s}$. 
The simulation samples used in the SME weights computation were generated according to differential SM cross sections, involving a convolution of the SM matrix element and the parton distribution function. 
Since the energy carried by the incoming particles relies on the parton distribution function at a given center-of-mass energy $\sqrt{s}$, the raising of $f(t)$ with $\sqrt{s}$ was investigated by generating dedicated samples turning off parton distribution functions in the proton, thus any remaining increase in amplitude could be attributed to the SME matrix element expression.
An enhancement of the $f(t)$ amplitude as the square of the center-of-mass energy was found with these samples, compatible with the expressions for the matrix elements given in \cite{Berger:2015yha}.

At Tevatron, $p\bar{p}\rightarrow t\bar{t}$ production was initiated mainly by $q\bar{q}$ annihilation while at the LHC, with the increase of $\sqrt{s}$, $gg$ fusion is dominant in $pp\rightarrow t\bar{t}$ production owing to higher gluon luminosity in the proton. 
We compared the $f(t)$ amplitudes obtained for $p\bar{p}$ collisions at D\O~and $pp$ collisions in the CMS laboratory frame, in samples generated at the same center-of-mass energy $\sqrt{s}=1.96$ TeV, and examine separately $gg$ and $q\bar{q}$ production mechanisms. 
We find relatively larger amplitude for $f(t)$ in $q\bar{q}$ and in $gg$ production mechanisms, and for a given $\sqrt{s}$, the experiment's position can favor one benchmark scenario or the other. 
Overall, the increase in center-of-mass-energy has a dominant impact on the amplitude, while the change in detector position and production mechanism induces a smaller change.

We scanned the latitude and azimuth of potential experiments on earth (the longitude does not impact the amplitude owing to the earth rotation). It was found that both ATLAS or CMS sit in a dip for the projected sensitivity on the benchmarks  $c_{XX} =-c_{YY} \neq 0$ and $c_{XY} =c_{YX} \neq 0$. In general, ATLAS and CMS sensitivity to SME coefficients will be similar since they are located at opposite azimuthal angle in the LHC ring.

\section{Sensitivity at the LHC and future colliders}

In this section, projected sensitivity at the LHC and future colliders will be studied and compared with Tevatron results. 
The D\O~analysis \cite{Abazov:2012iu} at Tevatron was performed with a luminosity of 5.3 $fb^{-1}$ of $p - \bar{p}$ collisions at $\sqrt{s}=1.96$ TeV. 
The LHC produces $p-p$ collisions, with about 150 $fb^{-1}$ of data recorded at $\sqrt{s}=13$ TeV during Run 2 \cite{Sirunyan:2019osb}. 
The High Luminosity LHC (HL-LHC), expected to start data taking in 2026, will deliver 3 $ab^{-1}$ at $\sqrt{s}=14$ TeV \cite{ApollinariG.:2017ojx}. 
The High Energy LHC (HE-LHC) is a future collider option that could take place after the HL-LHC, using the same tunnel with upgraded magnets, to achieve an expected center-of-mass energy of $\sqrt{s}=27$ TeV and 15 $ab^{-1}$ of integrated luminosity \cite{Zimmermann:2651305}. 
Eventually the Future Circular Collider, in its hadron collider stage (FCC-hh), is an even higher energy option, where a new 100 km tunnel nearby CERN would be built to achieve the unprecedented center-of-mass energy of $\sqrt{s}=100$ TeV and 15 $ab^{-1}$ of integrated luminosity \cite{Benedikt:2651300}. 
For simplicity, we assume the same coordinates as the CMS detector for an experiment at the LHC and future colliders (LHC Run 2, HL-LHC, HE-LHC, and FCC).

The SM cross section for $t\bar{t}$ production is computed at the center-of-mass energy of each collider scenario with Top++ \cite{Czakon:2011xx}, at next-to-next-to-leading order accuracy in perturbative QCD.

The hypothesis is made that the efficiencies for selecting reconstructed particles are identical to those of \cite{Khachatryan:2016kzg} at the LHC and beyond. 
This can be regarded as optimistic if considering the increasing number of collisions piling up with the hard process at higher and higher instantaneous luminosity (from about 30 pileup events at the LHC to 1000 at the FCC). However pileup mitigation techniques have been proved to work very efficiently, and more ideas are being explored to keep pileup impact under control at the future detectors \cite{Benedikt:2651300}.

We study the $e\mu$ final state, where the background is arising mainly from single top production, and  the Drell-Yan production is efficiently suppressed by requiring two leptons of different flavour. 
The same signal to background ratio as in \cite{Khachatryan:2016kzg} is also assumed (this $t\bar{t}$ channel is usually very clean with $s/b \approx 15$). This approximation could be refined by computing the cross section of the main backgrounds with fixed order QCD calculation, however the present value of the ratio is believed to be reasonably stable at higher center-of-mass energies.

The expected contributions for the LIV signal, SM $t\bar{t}$ production and single top background are fitted with a $\chi^2$ method to the Asimov dataset \cite{Cowan:2010js}, using bins of one sidereal hour. The Asimov dataset represents fake data constructed from the sum of all contributions excluding the signal. 
We perform the study for the above mentioned colliders and SME coefficient benchmarks. 
We obtain the projected precision on the SME coefficients. As a cross-check, we find sensitivity of the same order of magnitude as with $\chi^2$ method, when using $\mbox{\textsc{HistFactory}}$ \cite{Cranmer:2012sba} implementing the LHC test-statistics in a likelihood fit \cite{CMS-NOTE-2011-005}.  
Systematic uncertainties are rounded from \cite{Khachatryan:2016kzg}: 2\% is attributed to the luminosity, 4\% on the inclusive measurement of $t\bar{t}$ production, and 30\% on the small single top background.
Projected precision on the SME parameters is shown on table~\ref{tab:1}.

 \begin{table}[th!]
	\caption{Comparison of expected precision in the measurement of the SME parameters in $t\bar{t}$ signature for D\O, LHC Run 2, HL-LHC, HE-LHC, FCC experiment.}
	\label{tab:1}  
	\scalebox{0.73}{    \begin{tabular}{cccccc}
        \hline\noalign{\smallskip}
         &  D$\emptyset$  & LHC (Run II)  & HL-LHC  & HE-LHC & FCC\\
        \noalign{\smallskip}\hline\noalign{\smallskip}
        $\Delta c_{LXX} , \Delta c_{LXY}$ & $1\times 10^{-1}$  & $7\times 10^{-4}$ & $2\times 10^{-4}$ & $2\times 10^{-5}$ & $5\times 10^{-6}$ \\
        $\Delta c_{LXZ} , \Delta c_{LYZ}$ & $8\times 10^{-2}$ & $3\times 10^{-3}$ & $5\times 10^{-4}$ & $9\times 10^{-5}$ & $2\times 10^{-5}$ \\
        \noalign{\smallskip}\hline\noalign{\smallskip}
        $\Delta c_{RXX} , \Delta c_{RXY}$ & $9\times 10^{-2}$  & $3\times 10^{-3}$ & $5\times 10^{-4}$ & $8\times 10^{-5}$ & $5\times 10^{-5}$ \\
        $\Delta c_{RXZ} , \Delta c_{RYZ}$ & $7\times 10^{-2}$ & $1\times 10^{-2}$ & $2\times 10^{-3}$ & $4\times 10^{-4}$ & $8\times 10^{-5}$ \\
        \noalign{\smallskip}\hline\noalign{\smallskip}
        $\Delta c_{XX} , \Delta c_{XY}$ & $7\times 10^{-1}$  & $1\times 10^{-3}$ & $2\times 10^{-4}$ & $3\times 10^{-5}$ & $9\times 10^{-6}$ \\
        $\Delta c_{XZ} , \Delta c_{YZ}$ & $6\times 10^{-1}$ & $4\times 10^{-3}$ & $7\times 10^{-4}$ & $1\times 10^{-4}$ & $3\times 10^{-5}$ \\
        \noalign{\smallskip}\hline\noalign{\smallskip}
        $\Delta d_{XX} , \Delta d_{XY}$ & $1\times 10^{-1}$  & $6\times 10^{-4}$ & $1\times 10^{-4}$ & $2\times 10^{-5}$ & $8\times 10^{-6}$ \\
        $\Delta d_{XZ} , \Delta d_{YZ}$ & $7\times 10^{-2}$ & $2\times 10^{-3}$ & $4\times 10^{-4}$ & $8\times 10^{-5}$ & $2\times 10^{-5}$ \\
        \noalign{\smallskip}\hline\vspace{0.1cm}
	\end{tabular}}
\end{table}

The expected precision found by performing the likelihood fit using D\O~location, the value of $A_P$ and $A_F$ matrices in \cite{Whittington:2012wsa} and the total number of observed events quoted in \cite{Abazov:2012iu}, is found to be compatible with the observed results in D\O~analysis, with an absolute precision of the order of 10\%, thus validating the procedure. 

The precision on the SME coefficients is expected to be improved by up to three orders of magnitude from D\O~to the LHC Run II, depending on the coefficients.
An additional expected improvement is found at future hadron colliders, with up to two more orders of magnitude at the FCC. Overall, performing sidereal time analysis of $t\bar{t}$ production at present and future hadron colliders will greatly improve existing bounds on Lorentz-violating $c_{\mu\nu}$ coefficients for the top quark in the SME.

It has to be noted that parton distribution functions in the proton at 100 TeV are subject to high uncertainties at large momentum transfer \cite{Rojo:2016kwu}. 
The expected results are also subject to other approximations relative to the performance of future detectors, the treatment of pileup, and the cross sections for top quark processes. 
Although we consider that the adopted approximations are reasonable, results of this phenomenology study should mainly be considered as providing an order of magnitude for the sensitivity rather than a precise and definitive answer, that will be given by future experiments. 

The improvement found in the expected precision of the SME coefficients at the LHC and future colliders is explained by a combination of three factors: 
1) the increase in SM $t\bar{t}$ cross sections with $\sqrt{s}$ relative to Tevatron, 
2) the higher expected number of events produced in collisions with the greater volume of integrated luminosity, and 
3) the increase in the SME over SM matrix elements for $t\bar{t}$ production and decay with $\sqrt{s}$, leading to an increase of the amplitude of the function $f(t)$ in eq.~\ref{eqFt2}. 

The present analysis can be refined in several ways to improve sensitivity to LIV in the experiments. In addition to the $e\mu$ channel of $t\bar{t}$ decay, the same flavour dilepton channel and the lepton+jets channel could be used. Eventually, the $c_{\mu\nu}$ coefficients are modifying top quark kinematics, thus differential cross sections or multivariate analysis making use of kinematic $t\bar{t}$ observables could be used to improve sensitivity.

\section{Conclusions}

In this paper, we highlighted the physics potential of the LHC and future hadron colliders for LIV searches with $t\bar{t}$ production. 
Bounds on the top quark $c_{\mu\nu}$ coefficients in the SME can be improved by up to three orders of magnitude already at the LHC, and the total improvement is expected to reach five orders of magnitude at future colliders such as the FCC. 

Other proposed searches in the top sector \cite{Berger:2015yha} are targeting CPT violation at hadron colliders, by measuring the charge asymmetry between single top and antitop events as a function of sidereal time. 
This search is experimentally very challenging, 
and would deserve dedicated sensitivity studies, that are postponed to a later paper. 

Other LIV processes of interest would deserve detailed studies. The LHC is often thought of as a top factory, however the production of QCD and electroweak particles has also a very high cross section. By studying the production of QCD jets, $W^{\pm}$ and $Z$ bosons at present and future hadron colliders, poorly constrained areas of the SME could be probed at an unprecedented sensitivity.

\section*{Acknowledgements}
We would like to thank Alan Kosteleck\'y for enlightening discussions.

\bibliographystyle{spphys}       
\bibliography{PaperLIVcolliders}   

\end{document}